\documentclass[a4paper,fleqn,usenatbib,useAMS]{mnras}

\usepackage{graphicx}	
\usepackage{amsmath}	
\usepackage{amssymb}	
\usepackage{multicol}        
\usepackage{bm}		
\usepackage{pdflscape}	

\usepackage[T1]{fontenc}
\usepackage{ae,aecompl}

\title[New stellar system candidates]{
The Dark Energy Survey view of the Sagittarius stream: discovery of two faint stellar system candidates}

\author[E. Luque et al.]{\Large E. Luque,$^{1,2}$\thanks{\textbf{E-mail}: elmer.luque@ufrgs.br} A. Pieres,$^{1,2}$ B. Santiago,$^{1,2}$ B.~Yanny,$^3$ A.~K.~Vivas,$^4$ A.~Queiroz,$^{1,2}$ 
\newauthor\Large A.~Drlica-Wagner,$^{3}$ E.~Morganson,$^5$ E.~Balbinot,$^{6}$ J.~L.~Marshall,$^{7}$ T.~S.~Li,$^{7}$ A.~Fausti Neto,$^{2}$
\newauthor\Large  L.~N.~da Costa,$^{2,8}$ M.~A.~G.~Maia,$^{2,8}$ K.~Bechtol,$^{9}$ A.~G.~Kim,$^{10}$ G.~M.~Bernstein,$^{11}$  
\newauthor\Large S.~Dodelson,$^{3,9}$ L.~Whiteway,$^{12}$ H.~T.~Diehl,$^{3}$ D.~A.~Finley,$^{3}$ T.~Abbott,$^{4}$ F.~B.~Abdalla,$^{12,13}$ 
\newauthor\Large  S.~Allam,$^{3}$ J.~Annis,$^{3}$ A.~Benoit-L{\'e}vy,$^{12,14,15}$ E.~Bertin,$^{14,15}$ D.~Brooks,$^{12}$ D.~L.~Burke,$^{16,17}$
\newauthor\Large  A. Carnero Rosell,$^{2,8}$ M.~Carrasco~Kind,$^{5,18}$ J.~Carretero,$^{19,20}$ C.~E.~Cunha,$^{16}$ 
\newauthor\Large C.~B.~D'Andrea,$^{21,22}$ S.~Desai,$^{23}$ P.~Doel,$^{12}$ A.~E.~Evrard,$^{24,25}$ B.~Flaugher,$^{3}$ P.~Fosalba,$^{19}$ 
\newauthor\Large D.~W.~Gerdes,$^{25}$ D.~A.~Goldstein,$^{10,26}$  D.~Gruen,$^{16,17}$ R.~A.~Gruendl,$^{5,18}$ G.~Gutierrez,$^{3}$ 
\newauthor\Large D.~J.~James,$^{4}$  K.~Kuehn,$^{27}$ N.~Kuropatkin,$^{3}$ O.~Lahav,$^{12}$ P.~Martini,$^{28,29}$ R.~Miquel,$^{20,30}$
\newauthor\Large  B.~Nord,$^{3}$ R.~Ogando,$^{2,8}$ A.~A.~Plazas,$^{31}$ A.~K.~Romer,$^{32}$ E.~Sanchez,$^{33}$ V.~Scarpine,$^{3}$    
\newauthor\Large  M.~Schubnell,$^{25}$ I.~Sevilla-Noarbe,$^{33}$ R.~C.~Smith,$^{4}$ M.~Soares-Santos,$^{3}$ F.~Sobreira,$^{2,34}$      
\newauthor\Large E.~Suchyta,$^{35}$ M.~E.~C.~Swanson,$^{5}$ G.~Tarle,$^{25}$ D.~Thomas$^{21}$ and A.~R.~Walker$^{4}$
\vspace*{1em}\\\noindent
Affiliations are listed at the end of the paper
}

\date{Released \today}

\pubyear{2015}

\begin{document}
\label{firstpage}
\pagerange{\pageref{firstpage}--\pageref{lastpage}}
\maketitle

\begin{abstract}
We report the discovery of two new candidate stellar systems in the constellation of Cetus using the data from the first two years of the Dark Energy Survey (DES). The objects, DES\,J0111$-$1341 and DES\,J0225$+$0304, are located at a heliocentric distance of $\sim 25\,\mathrm{kpc}$ and appear to have old and metal-poor populations. Their distances to the Sagittarius orbital plane, $\sim 1.73\,\mathrm{kpc}$ (DES\,J0111$-$1341) and $\sim 0.50\,\mathrm{kpc}$ (DES\,J0225$+$0304), indicate that they are possibly associated with the Sagittarius dwarf stream. The half-light radius ($r_\mathrm{h}\simeq 4.55\,\mathrm{pc}$) and luminosity ($M_V\simeq +0.3$) of DES\,J0111$-$1341 are consistent with it being an ultrafaint stellar cluster, while the half-light radius ($r_\mathrm{h}\simeq 18.55\,\mathrm{pc}$) and luminosity ($M_V\simeq -1.1$) of DES\,J0225$+$0304 place it in an ambiguous region of size--luminosity space between stellar clusters and dwarf galaxies. Determinations of the characteristic parameters of the Sagittarius stream, metallicity spread ($-2.18 \lesssim\mathrm{[Fe/H]} \lesssim -0.95$) and distance gradient ($23\,\mathrm{kpc}\lesssim \mathrm{D}_{\sun} \lesssim 29\,\mathrm{kpc}$), within the DES footprint in the Southern hemisphere, using the same DES data, also indicate a possible association between these systems. If these objects are confirmed through spectroscopic follow-up to be gravitationally bound systems and to share a Galactic trajectory with the Sagittarius stream, DES\,J0111$-$1341 and DES\,J0225$+$0304  would be the first ultrafaint stellar systems associated with the Sagittarius stream. Furthermore, DES\,J0225$+$0304 would also be the first confirmed case of an ultrafaint satellite of a satellite.
\end{abstract}

\begin{keywords}
globular cluster: general -- galaxies:dwarf.
\end{keywords}




\section{Introduction}
\label{sec:intro}
The Sagittarius dwarf galaxy was discovered relatively recently due to its position on the far side of the Milky Way \citep[MW;][]{Ibata1994}. Its extended morphology towards the MW plane suggested the existence of extra tidal features \citep{Johnston1995,Lynden-Bell1995,Mateo1996}. The Two Micron All-Sky Survey (2MASS) and the Sloan Digital Sky Survey (SDSS) made it clear that this dwarf is responsible for the most conspicuous tidal stellar substructure present in the Galactic halo \citep{Newberg2002,Majewski2003}.

Deeper photometric and spectroscopic data, specifically with SDSS, have allowed the morphological, structural and kinematic properties of the Sagittarius stream to be disentangled from MW substructure \citep{Newberg2003,Belokurov2006,Belokurov2014,Newberg2007,Yanny2009}. This wealth of data was used by \citet{Law2010} to model the MW gravitational potential and to find some evidence in favour of triaxiality (i.e., flattening).

\begin{figure*} 
\includegraphics[width=1.\textwidth]{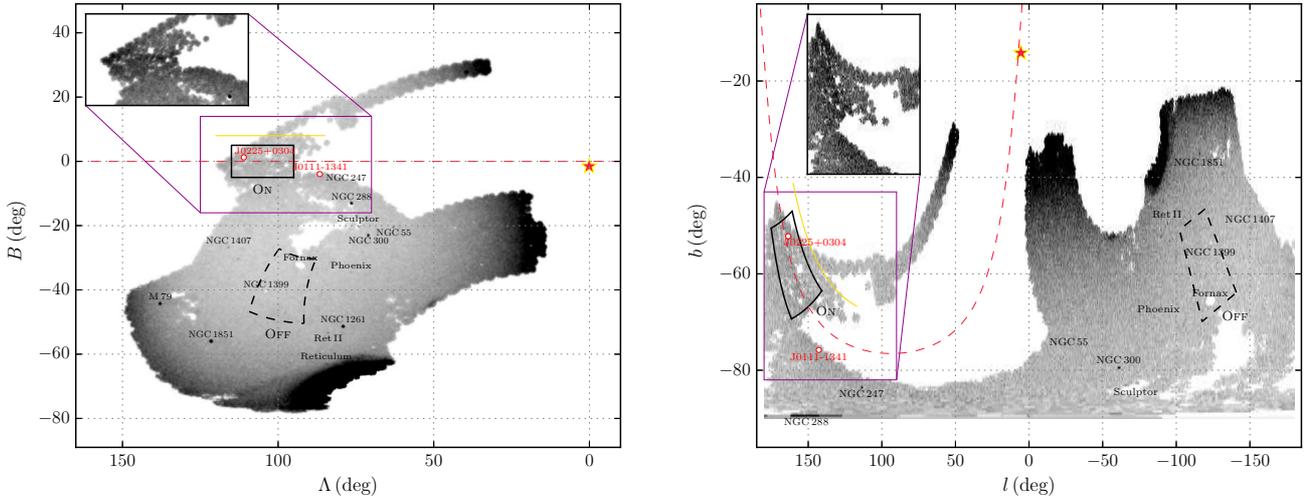}
\caption{Density map of stars with $17<g<23$ and $-0.5<g-r<1.2$ from the DES Y2Q1 footprint in two different coordinate systems. The left-hand panel is shown in a coordinate system defined by the orbit of the Sagittarius dwarf \citep{Majewski2003, Belokurov2014}, while the right-hand panel is in Galactic coordinates ($l,b$). The current location of the Sagittarius dwarf is indicated by a red star. The red dashed line traces the Sagittarius tidal tail \citep{Majewski2003}. At the top left of each panel, we show an inset of the density map where the stream is more visible.  Overdensities of some GCs \citep{Harris2010} and dwarf galaxies \citep{McConnachie2012,Bechtol2015,Koposov2015} are also indicated. The new stellar system candidates discovered, DES\,J0111-1341 and DES\,J0225+0304, are marked with red circles. The {\sc On} region (solid lines) defined by $95\degr<\Lambda<115\degr$ and $-5\degr<B<5\degr$ represents the best-sampled region of the Sagittarius stream, whereas the {\sc Off} region (dashed lines) represents the region of stars used for the background with the same Galactic latitude as the {\sc On} region. The Fornax dwarf galaxy and NGC\,1399 were masked to avoid overestimating the density of our sample of background stars (\textsc{Off} region). These regions are used to construct the Hess diagrams in Figs \ref{fig:Hess_sgt} and \ref{fig:Hess_along}. The yellow solid line, as shown in both panels, indicates the position of the faint branch of the Sagittarius stream in the DES footprint (see Section \ref{subsec:Counting stars}).  
}
\label{fig:maps_Sg}
\end{figure*} 

\citet{Belokurov2006} demonstrated that the Sagittarius stream in the northern Galactic hemisphere bifurcates into brighter and fainter components separated by up to $\sim 15\degr$ on the sky. More recently, \citet{Koposov2012} have shown  that a bifurcation also appears in the Sagittarius tails in the southern Galactic hemisphere. This fainter branch had comparatively more metal-poor stars and a simpler mix of stellar populations than the main structure. The southern bifurcation, extending at least $30\degr$ on the sky, was confirmed using Panoramic Survey Telescope and Rapid Response System 1 (Pan-STARRS) data by \citet{Slater2013}. The authors found evidence that the fainter substructure is $5\,\mathrm{kpc}$ closer to the Sun than the brighter one, similar to the behaviour seen in the northern Galactic hemisphere. They also argue that the distance between the streams agrees with the predictions of the $N$--body simulations presented by \citet{Law2010}. Based on their model, the same authors also identify MW satellites, dwarf galaxies and globular clusters (GCs), that may be physically associated with the Sagittarius dwarf. In particular, the Sagittarius dwarf has been observed to contain at least four  GCs (NGC\,6715, Arp\,2, Terzan\,7, and Terzan\,8) within its main body \citep{Dacosta1995,Bellazzini2003}. However, different studies have proposed several GCs to likely be associated with the Sagittarius stream \citep[e.g.,][]{Bellazzini2003,Forbes2010,Dotter2010,Dotter2011,Carballo2014,Sbordone2015}. Even open clusters (OCs) have been suggested as members of the Sagittarius family \citep[e.g.,][]{Carraro2004,Carraro2009}.
It is likely that additional GCs and OCs may have been stripped from Sagittarius during prolonged interaction with the MW and now lie scattered throughout the Galactic halo. In a recent analysis based on new models of the tidal disruption of the Sagittarius dwarf, \citet{LawMajewski2010} found that several of the candidates proposed in the literature have non-negligible probability of belonging to the Sagittarius dwarf. However, calculating the expected quantity of false associations in the sample, they proposed that only five GCs (Arp\,2, NGC\,6715, NGC\,5634, Terzan\,8, and Whiting\,1) are almost certainly associated with the Sagittarius dwarf, an additional four (Berkeley\,29, NGC\,5053, Pal\,12, and Terzan\,7) are moderately likely to be associated.

It now appears that stars left over from the accretion of the Sagittarius dwarf entirely wrap around the Galactic centre. Recent spectroscopic analysis by \citet{Hyde2015}, for instance, finds over 100 good stream candidates with metallicities in the range $-0.97<[\mathrm{Fe/H}]< -0.59$ spread over $142\degr$. \citet{deBoer2015} analyse the stream in the SDSS Stripe 82 region with both photometry and spectroscopy, finding a tight age--metallicity relation. They also show that the fainter branch is old ($> 9\,\mathrm{Gyr}$) and metal-poor ($[\mathrm{Fe/H}] < -1.3$), while the dominant branch has stars covering large ranges in age and metallicity.
 
In this paper we explore the tidal tails of Sagittarius within the Dark Energy Survey (DES; \citealt{DES2005}) footprint in the Southern hemisphere. This data set is $\sim 2\,\mathrm{mag}$ (in the $g$ band) deeper than other large surveys covering this part of the sky (e.g., Pan-STARRS or SDSS). DES is a wide-field imaging survey of the Southern hemisphere that has recently finished its third year of data taking, from an expected total of 5\,yr \citep{Diehl2016}. We also identify two previously undiscovered ultrafaint stellar systems whose inferred ages, metallicities and distances make it likely that they are associated with Sagittarius. In Section \ref{sec:data}, we present the DES data. In Section \ref{sec:Sgt}, we discuss the properties of the Sagittarius stream as probed by those data. In Section \ref{sec:method}, we present a method used to search for star clusters and other stellar systems in the DES footprint. In Section \ref{sec:J0111_J0225}, we report on the identification of the two star system candidates whose properties make them likely to have been stripped from the Sagittarius dwarf. If DES\,J0111$-$1341 is confirmed to be a stellar cluster, it will be named DES\,2, whereas DES\,J0225$+$0304 will be named Cetus\,III if found to be a dwarf galaxy. Our final remarks are then presented in Section \ref{sec:conclusions}.

\section{DES Data}
\label{sec:data}
DES is a wide-field optical imaging survey of $~5000\,\mathrm{deg}^2$ in the southern equatorial hemisphere in the \textit{grizY} bands. DES is scheduled for 525 nights distributed over 5\,yr. It uses the Dark Energy Camera (DECam), a 3 $\deg^2$ ($2\fdg2$ diameter) mosaic  camera with $0.263\,\mathrm{arcsec}$ pixels on the prime focus of the 4-metre Blanco telescope at Cerro Tololo Inter-American Observatory \citep{Flaugher2015}. The DES data management (DESDM) uses an image processing pipeline that consists of image detrending, astrometric calibration, nightly photometric calibration, global calibration, image coaddition, and object catalogue creation. For a more detailed description, we refer to  \citet{Sevilla2011}, \citet{ Desai2012} and \citet{Mohr2012}. Here, we use DES Y2Q1 (year-two quick release) data derived from single-epoch imaging. This catalogue is derived from 26,590 DECam exposures taken during the first 2\,yr of DES observing and has a median point-source depth at an $\mathrm{S/N}=10$ of $g=23.4$, $r=23.2$, $i=22.4$, $z=22.1$, and $Y=20.7$. The resulting calibrated DES magnitudes are already corrected for Galactic reddening by the stellar locus regression (SLR) calibration \citep[see][]{Drlica2015}.

The stellar sample used in this work was drawn using {\sc sextractor} parameters ${\tt FLAGS}$, ${\tt SPREAD\_MODEL}$ and ${\tt PSF}$ magnitudes \citep{Bertin1996,Bertin2011,Bouy2013}. Briefly, ${\tt FLAGS}$ tells for instance if an object is saturated or has been truncated at the edge of the image, while ${\tt SPREAD\_MODEL}$ is the main star/galaxy separator. We used the weighted-average (${\tt WAVG}$) of the ${\tt SPREAD\_MODEL}$ measurements from the individual exposures of each source. A source quality criterion of ${\tt FLAGS}<4$ over the $g$ and $r$ filters was also applied. To increase stellar completeness, we selected sources in the $r$-band with ${\tt |WAVG\_SPREAD\_MODEL|}< 0.003+ {\tt SPREADERR\_MODEL}$. 
A bright (faint) $g$ magnitude limit of ${\tt WAVG\_MAG\_PSF}=17$ (${\tt WAVG\_MAG\_PSF}=24$) was also applied. In order to prevent point sources with extreme colours (including red dwarfs from the Galactic disc) from contaminating the sample, a colour cut at $-0.5 \leq g-r\leq 1.2$ was also used. \citet{Drlica2015} show that our Y2Q1 stellar classification efficiency exceeds $90\%$ for $r<23\,\mathrm{mag}$, and  falls to $\sim 80\%$ by $r<24\,\mathrm{mag}$, whereas contamination by galaxies is $\sim 40\%$ for $23\,\mathrm{mag}$ and increases to $\sim 60\%$ by $r<24\,\mathrm{mag}$.  

\begin{figure*}\centering
\includegraphics[width=1.\textwidth]{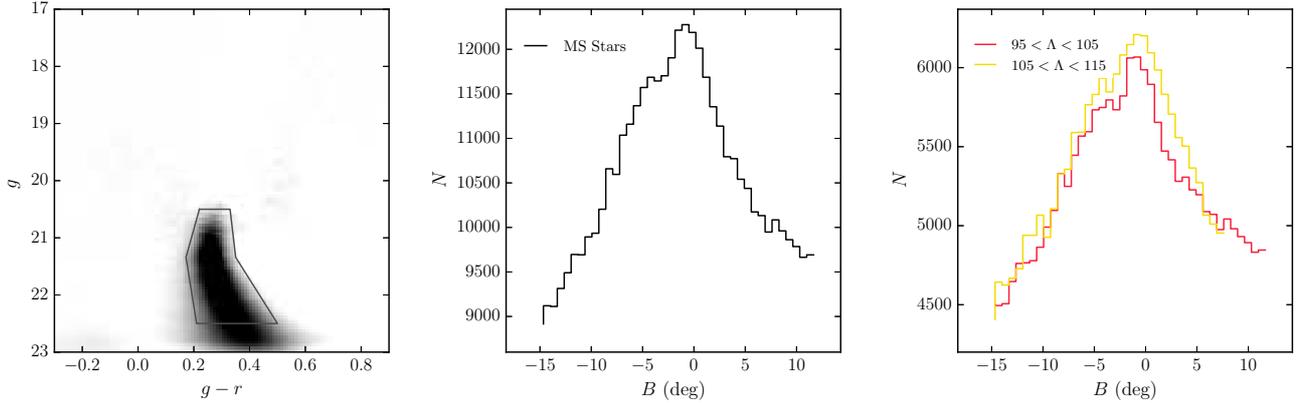}
\caption{Left-hand panel: decontaminated Hess diagram constructed with stars within a region defined by $95\degr <\Lambda< 115\degr$ and $-15\degr<B<12\degr$. The solid lines on the CMD plane are used to select MS stars belonging to the Sagittarius stream. Middle panel: number of MS stars (using the selected stars in the previous panel) along the Sagittarius stream ($95\degr <\Lambda< 115\degr$). The expected position of the faint stream (\citealt{Koposov2012}) is at $B \sim 8\degr$. Right-hand panel: number of MS stars for two regions along the stream from $95\degr  < \Lambda < 105\degr$ and $105\degr  < \Lambda < 115\degr$  as indicated. The faint stream is seen only at $95\degr  < \Lambda < 105\degr$ due to the current coverage of the DES footprint (see Fig. \ref{fig:maps_Sg}).}\label{fig:hist_sgt}
\end{figure*}

\section{Sagittarius stream in the southern hemisphere}
\label{sec:Sgt}
In Fig. \ref{fig:maps_Sg}, we show a density map of stars with $17<g<23$ and $-0.5<g-r<1.2$ from the DES Y2Q1 footprint in two different coordinate systems. The colour cut was performed to exclude stars from the Galactic disc and possibly spurious objects that can contaminate our sample. The left-hand panel is in the coordinate system aligned with the Sagittarius stream ($\Lambda,B$)  \citep{Majewski2003,Belokurov2014}, while the right-hand panel is in Galactic coordinates ($l,b$). Several overdensities are noticeable, such as some GCs \citep{Harris2010}, dwarf galaxies \citep{McConnachie2012} and the recently discovered dwarf galaxy Reticulum\,II \citep{Bechtol2015,Koposov2015}. The Sagittarius stream in the Southern hemisphere (trailing tail) is also visible between $90\degr\lesssim\Lambda\lesssim 120\degr$ and $-15\degr\lesssim B\lesssim 12\degr$ in Sagittarius coordinates and between $120\degr\lesssim l\lesssim 190\degr$ and $-80\degr\lesssim b\lesssim -45\degr$ in Galactic coordinates (see the inset maps on the top left of each panel). In the same figure, we show with red circles two new stellar system candidates, DES\,J0111$-$1341 and DES\,J0225$+$0304. Given their physical locations, these new candidates are possibly associated with the Sagittarius stream (discussed in Section \ref{sec:method}). In this figure, we also show \textsc{On} and \textsc{Off} regions. The \textsc{On} region (solid lines) defined by $95\degr\leq\Lambda\leq 115\degr$ and $-5\degr\leq B\leq 5\degr$ represents the best sampled region of the Sagittarius stream, while the \textsc{Off} region (dashed lines) represents the sample of background\footnote{We refer to these stars as `background', though they are dominantly composed of MW foreground stars.} stars located at the same Galactic latitude as the \textsc{On} region. These regions are used in our colour-magnitude diagram (CMD) analysis presented in Sections \ref{sec:spread_met} and \ref{sub:cmd_analysis}. Finally, the yellow solid line represents the position of a possible secondary peak previously identified by \citet[][see discussion in the next section]{Koposov2012}.

We emphasize that our analysis of the Sagittarius stream is focused on determining, using DES data, its basic characteristic parameters, such as metallicity, age and distance ranges, so that they can be compared to the properties inferred for the newly discovered systems, DES\,J0111$-$1341 and DES\,J0225$+$0304. The compatibility between stream stars and these newly found systems helps shed light on their possible physical association. 

\subsection{Inferred Number of Stars}
\label{subsec:Counting stars}
The Sagittarius stream is known to display substructures, like its bright and faint branches, both in the northern and southern Galactic hemispheres \citep{Newberg2003,Belokurov2006,Yanny2009,Koposov2012}. In particular, in the southern Galactic hemisphere, parallel to the bright branch, but $\sim 10\degr$ away, the faint branch is found \citep{Koposov2012}. We start by analysing variations in stellar number counts along and across the Sagittarius stream as covered by DES, in search for any clear branching of the stream in this region. In the left-hand panel of Fig. \ref{fig:maps_Sg}, we show the density map of the Sagittarius stream in the coordinate system approximately aligned with the orbit of Sagittarius $(\Lambda, B)$ as described in \citet{Majewski2003} and \citet{Belokurov2014}. We selected stars inside an area defined by $95\degr<\Lambda<115\degr$ and $-15\degr<B<12\degr$. We name this region the \textit{stream sample}. This chosen region is a compromise between reaching a reasonably homogeneous stream coverage along both streams and still keeping a sizeable area within the DES footprint. To subtract the expected number of background stars coinciding with the
\textit{stream sample} region, we selected stars inside a region that is offset by $\Delta l=80\degr$ with respect to the centre of the \textit{stream sample} region. These regions\footnote{These two regions are not shown in Fig. \ref{fig:maps_Sg} to avoid confusion with the \textsc{On} and \textsc{Off} regions used in Sections \ref{sec:spread_met} and \ref{sub:cmd_analysis}.} have approximately the same area and are from approximately the same Galactic latitude ($b\sim -59\degr$) in order to maintain similar background density. For each region described above, we constructed the Hess diagram. In the left-hand panel of Fig. \ref{fig:hist_sgt}, we show the decontaminated Hess diagram calculated as the difference between the Hess diagrams of both regions weighted by their respective areas\footnote{We replace negative values in the decontaminated Hess diagram by zero.}. We use the {\sc healpix} software to determine the effective area in each region. In order to obtain a sample of representative stars of the Sagittarius stream, we weight each star of the \textit{stream sample} region by its probability of being member of the Sagittarius stream, $w = n_i/m_i$, where $n_i$ ($m_i$) represents the number of stars in a given cell of the Hess diagram, with bins of $0.01\,\mathrm{mag} \times 0.05\,\mathrm{mag}$, after (before) subtracting the background stars. We consider that all the stars in a given cell of the Hess diagram have the same weight. The solid lines in the CMD plane on the left-hand panel of Fig. \ref{fig:hist_sgt} select main sequence (MS) stars associated with the stream. We then use the weights of these stars to analyse the variation of the number of stars along and across the stream. The results are shown in the middle and right-hand panels of Fig. \ref{fig:hist_sgt}. We use the {\sc healpix} software to compute the area actually covered by the Y2Q1 footprint and thus to compensate the number of stars for the area loss.  

\citet{Koposov2012} find evidence for a faint stream at $B\sim 8\degr-10\degr$. The DES footprint covers this area only from $95\degr < \Lambda < 105\degr$. The red histogram in the right-hand plot in Fig. \ref{fig:hist_sgt} shows the number of MS stars within this region in bins of $B$; within this area, our data show a suggestion of an excess of stars that could be attributed to the faint stream. At Sagittarius longitudes $105\degr < \Lambda < 115\degr$, DES does not cover the secondary stream ($5\degr < B <12\degr$); however, where DES has coverage ($-15\degr < B <5\degr$), the number of MS stars (yellow histogram) is consistent with those at $95\degr < \Lambda < 105\degr$ (red histogram). Scaling the number of stars to the full range ($-15\degr < B <12\degr$), we infer a number of MS stars for $95\degr < \Lambda <115\degr$ as shown in the middle panel. The possible excess of stars observed at $B\sim 8\degr$ (middle and right-hand panels in Fig. \ref{fig:hist_sgt}) only is visible when we use bin sizes between $0\fdg 6 \lesssim \Delta B \lesssim 0\fdg 7$, otherwise, this latter is not evident. Therefore, in this paper, we do not claim a detection of the branching of the stream.

\begin{figure*}\centering
\includegraphics[width=.99\textwidth]{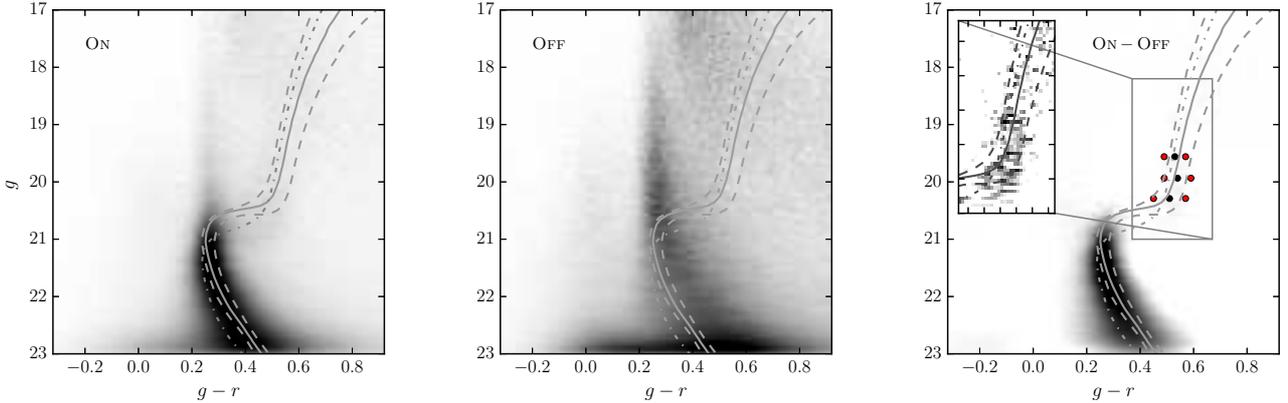}
\caption{Left-hand panel: Hess diagram constructed with stars within a region of the Sagittarius stream defined by $95\degr<\Lambda<115\degr$ and $-5\degr<B<5\degr$ (\textsc{On} region; Fig. \ref{fig:maps_Sg}). Middle panel: Hess diagram of the background constructed with stars within the \textsc{Off} region (Fig. \ref{fig:maps_Sg}). Right-hand panel: difference Hess diagram between the \textsc{On} and \textsc{Off} regions. The solid black (red) circles represent the mean colour (standard deviation) values of RGB stars as a function of the colour ($g-r$) for different ranges of magnitude. The mean colour value ($\mu$) and standard deviation ($\sigma$) obtained from the Gaussian fit are: $\mu=0.527\pm 0.002$ and $\sigma=0.043\pm 0.002$ (top circles), $\mu=0.544\pm 0.003$ and $\sigma=0.051\pm 0.005$ (middle circles), $\mu=0.509\pm 0.004$ and $\sigma=0.061\pm 0.006$ (lower circles). The solid line, in each panel, represents the best-fitting isochrone \citep{Bressan2012} determined from mean colour values. The lower (top) dashed line represents the $1\sigma$ ($-1\sigma$) isochrone. From the bottom up, the isochrone parameters are: $\log(\mathrm{Age})=9.98$, $\mathrm{D}_{\sun}=24.5\,\mathrm{kpc}$ and $\mathrm{[Fe/H]}=-0.95$ (lower dashed line), $\log(\mathrm{Age})=10.02$, $\mathrm{D}_{\sun}=24.5\,\mathrm{kpc}$ and  $\mathrm{[Fe/H]}=-1.34$ (solid line), $\log(\mathrm{Age})=10.12$, $\mathrm{D}_{\sun}=24.5\,\mathrm{kpc}$ and $\mathrm{[Fe/H]}=-2.18$ (top dashed line). In addition, an isochrone model with $\mathrm{[Fe/H]}=-2.18$, $\log(\mathrm{Age})=10.12$ and $(m-M)_0 =17.31$ parameters is also overplotted on each panel (dot--dashed line; see Sections \ref{sec:spread_met} and \ref{sub:cmd_analysis}).
}\label{fig:Hess_sgt}
\end{figure*}  

\subsection{Metallicity spread}
\label{sec:spread_met}
The Sagittarius stream in the northern Galactic hemisphere and the celestial equator (Stripe\,82) is known for having a metallicity range \citep[e.g.,][]{Koposov2012,deBoer2015,Hyde2015}. In particular, using photometric and spectroscopic data within the SDSS Stripe\,82 region (region in common with the DES footprint), \citet{Koposov2012} determined that the stars belonging to the bright and faint branches cover a metallicity range from  $-2\lesssim[\mathrm{Fe/H}]\lesssim 0$, while \citet{deBoer2015} determined a metallicity range from $-2.5\lesssim[\mathrm{Fe/H}]\lesssim -0.3$. However, the brighter branch contains substantial numbers of metal-rich stars as compared to the fainter branch \citep{Koposov2012}. 

We now turn to a global analysis of the stellar populations contributing to the Sagittarius stream. We first use the red-giant branch (RGB) stars to find a spread in metallicity, as follows. First, we selected stars inside a region defined by $95\degr<\Lambda<115\degr$ and $-5\degr<B<5\degr$ (\textsc{On} region; left-hand panel of Fig. \ref{fig:maps_Sg}). The more restricted range in $B$ is meant to further reduce sky coverage effects and to avoid any possible contamination by the faint branch. Using these stars, we have constructed and decontaminated a Hess diagram representative of the Sagittarius stream. The left-hand and middle panels of Fig.  \ref{fig:Hess_sgt} show the Hess diagrams for the \textsc{On} and \textsc{Off} regions, respectively. They contain $185\,558$ and $117\,860$ stars (within an isochrone filter\footnote{The isochrone filter is constructed by using the best-fitting isochrone determined for mean colour values (see Fig. \ref{fig:Hess_sgt}). For isochrone filter details, we refer to \citet{Luque2016}.}), respectively. These regions are from approximately the same Galactic latitude (see right-hand panel of Fig. \ref{fig:maps_Sg}). The decontaminated Hess diagram shown in the right-hand panel of Fig. \ref{fig:Hess_sgt} was calculated as the difference between the Hess diagrams of the  \textsc{On} and \textsc{Off} regions weighted by their respective areas. It contains a total of $87\,810$ stars. We use the {\sc healpix} software to determine the effective area in each region. In the  decontaminated Hess diagram we can identify MS, RGB, and some younger population stars. 

We select stars within the decontaminated CMD region defined by $0.4<g-r<0.8$ and $19.3<g<20.5$. For each interval of $\sim 0.4\,\mathrm{mag}$ along the CMD, we count stars as a function of colour and use \textsc{python} package scipy.optimize\footnote{http://docs.scipy.org/doc/scipy-0.17.0/reference/optimize.html} to fit a  Gaussian distribution to determine the mean colour value and the associated standard deviation. The peak and $1\sigma$ deviations from it are shown as the black and red dots in the right-hand panel of Fig. \ref{fig:Hess_sgt}. We then choose a set of {\sc parsec} isochrone \citep{Bressan2012} models that visually agree with the RGB mean and associated $\pm 1\sigma$ colours resulting from the Gaussian fits as well as the observed main sequence turn-off (MSTO) and MS loci. This is done by imposing the following restrictions to the isochrones: (i) the model age and metallicity must respect the tight age--metallicity relation by \citet{deBoer2015} and (ii) a single distance must be used for the three sets of points along the RGB, MS and MSTO loci.

\begin{figure*}
\includegraphics[width=.99\textwidth]{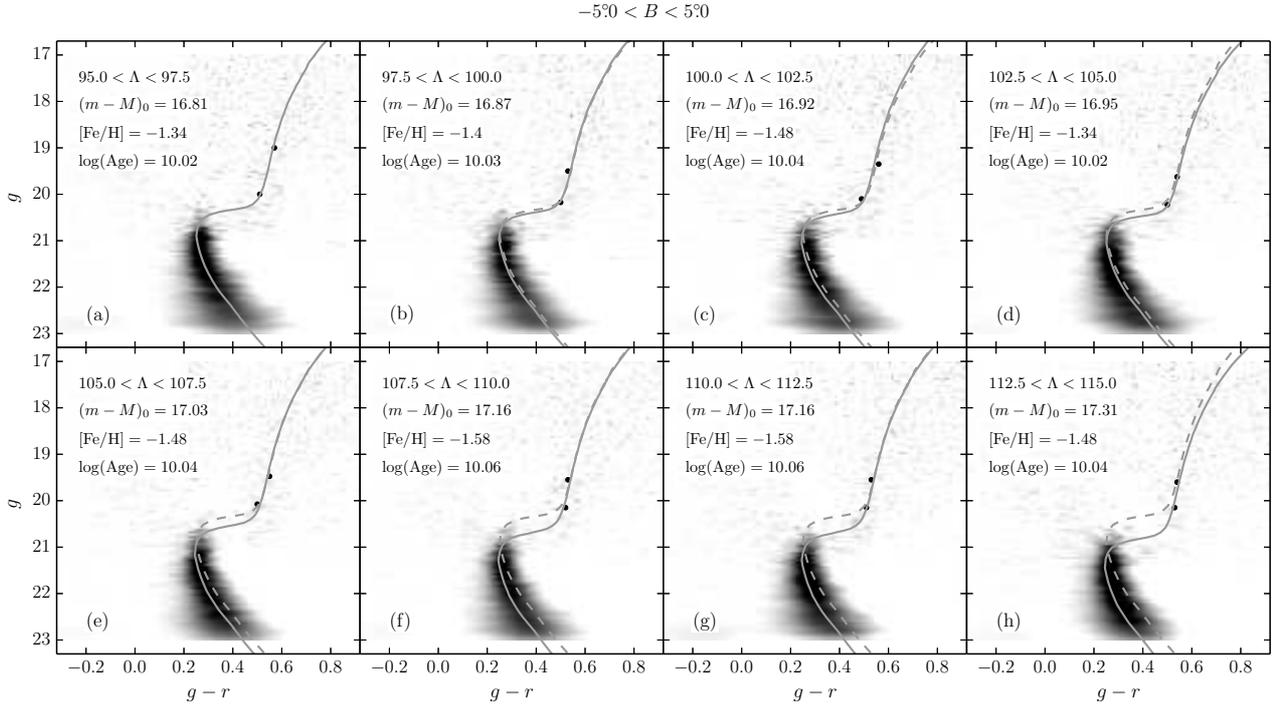}
\caption{Hess diagrams, after decontamination from the background contribution, constructed along the Sagittarius stream (\textsc{On} region) from $95\degr <\Lambda<97\fdg 5$ to $112\fdg 5<\Lambda< 115\degr$ as indicated on each panel. The solid black dots represent the mean colour values of the Gaussian fit on RGB stars. The best-fitting isochrone \citep{Bressan2012} using the method described in Section \ref{sub:cmd_analysis} is overplotted on each panel with solid line. The isochrone parameters are also indicated on each panel. For comparison, in panels (b) through (h) is overplotted (with a dashed line) the nearest isochrone model [from panel (a)].}
\label{fig:Hess_along}
\end{figure*}

The best-fitting isochrones for the mean values and standard deviations (as described above) are shown in Fig. \ref{fig:Hess_sgt}. Our results show that the stream population is old but displays a significant metallicity spread. While the peak RGB locus is consistent with $\mathrm{[Fe/H]}=-1.34$, their redder and bluer ends are more metal-rich ($\mathrm{[Fe/H]}=-0.98$) and metal-poor ($\mathrm{[Fe/H]}=-2.18$), respectively. The metallicity spread found in this analysis is also much larger than the photometric errors ($\sigma_{g-r}\simeq 0.01$ for RGB stars at $g\simeq 20$) and uncertainty in calibration\footnote{The uncertainty in calibration was determined by comparing the SLR calibration for Y2Q1 against external catalogues (2MASS and AAVSO Photometric All-Sky Survey, APASS-DR9). The latter transformed to DES filters.} [$\Delta(g-r)=0.013$ for the \textsc{On} region], which again attests to its reality. However, metallicity determinations in the literature \citep{Koposov2012,deBoer2015} suggest that the Sagittarius stream in the Stripe\,82 region contains more metal-rich stars than our determinations.

\subsection{Distance gradient}
\label{sub:cmd_analysis}
Distance determinations for different regions of the Sagittarius stream in the northern Galactic hemisphere were performed by different authors \citep[e.g,][]{Belokurov2006,Correnti2010}. 
Recent studies of the Sagittarius stream in the southern Galactic hemisphere were performed by \citet{Koposov2012} and \citet{Slater2013}. Using SDSS Data Release 8, \citet{Koposov2012} determined a distance gradient from $22.08$ ($\Lambda\simeq 97\fdg 5$) to $27.2\,\mathrm{kpc}$ ($\Lambda\simeq 112\fdg 5$), whereas \citet{Slater2013}, using Pan-STARRS data, determined a distance gradient from $29.5$ ($\Lambda\simeq 102\degr$) to $33.1\,\mathrm{kpc}$ ($\Lambda\simeq 110\degr$). We note a discrepancy in determining the distance gradient between the two groups. While it is true that both groups use red clump (RC) stars to determine the distance gradient along of the Sagittarius stream, the difference lies in the absolute magnitude value assumed in these determinations. To compare our results with the literature, we show only results within our region of analysis, $95\degr <\Lambda < 115\degr$. In this section, we perform an independent estimate of the distance gradient along the Sagittarius stream in the southern Galactic hemisphere, so as to compare it to those previous studies. 
  
For each interval of $2\fdg 5$ in $\Lambda$, we construct a Hess diagram along the Sagittarius stream (\textsc {On} region; left-hand panel of Fig. \ref{fig:maps_Sg}), eight in total. To decontaminate each one of the Hess diagrams by removing the background stars, we first divide the \textsc{Off} region in subregions approximately equal to those used in \textsc{On} region, maintaining the same Galactic latitude. We then follow the same procedure described in Section \ref{sec:spread_met}. The results are shown in Fig. \ref{fig:Hess_along}. 

We estimated the distance gradient along the stream as follows. For each interval in $\Lambda$ (see the text above), we first select CMD stars with $0.4<g-r<0.8$. For two intervals of magnitude, we count stars as a function of colour and fit a Gaussian distribution to determine the mean colour value ($\mu$). The choice of only two intervals of magnitude is due to low statistics of RGB stars present in all the CMDs.  By applying this restriction, we obtain  $\sim 150$ stars in each magnitude interval.  The peak values are shown as black solid dots in Fig. \ref{fig:Hess_along}. We then use a set of \textsc{parsec} isochrone \citep{Bressan2012} models that visually fit the RGB mean values resulting from the Gaussian fits as well as the observed MSTO and MS loci. This is done by again imposing the restriction that the model age and metallicity respect the  age--metallicity relation by \citet{deBoer2015}. The best-fitting isochrones using the method described above are superposed on to the decontaminated Hess diagrams (Fig. \ref{fig:Hess_along}). Our best-fitting  isochrones show a distance gradient along the Sagittarius stream from $\sim 23\,\mathrm{kpc}$ for $95\degr<\Lambda<97\fdg 5$ and $\sim 29\,\mathrm{kpc}$ for $112\fdg 5<\Lambda<115\degr$. We estimate distance uncertainties in each $\Lambda$ interval by varying the age and metallicity of the \textsc{parsec} isochrones around the best fitting case (but still bound to the same age--metallicity relation) and redoing the visual isochrone fit. We estimate a mean distance  uncertainty of $\pm 0.3\,\mathrm{kpc}$. Therefore, our results are in agreement with those obtained by \citet{Koposov2012}.  

To quantify the effect of the distance gradient  along the stream on the metallicity spread (see Section \ref{sec:spread_met}), we overplotted in Fig. \ref{fig:Hess_sgt} the same isochrone model that best fits the $-1\sigma$ Gaussian locus ($\mathrm{[Fe/H]}=-2.18$, $\log(\mathrm{Age})=10.12$; dots shown on the blue side of the RGB locus), but now shifted to $(m-M)_0 =17.31 = 29\,\mathrm{kpc}$ (dot--dashed line). This distance corresponds to the maximum value determined in the analysis of the distance gradient. We conclude that a variation in distance as large as is inferred in this section does not account for the observed colour spread on the RGB (see Fig. \ref{fig:Hess_sgt}). 

\section{Substructure Search and Object Detection}
\label{sec:method}
Many more than 17 objects were selected by our compact overdensity search techniques, stellar density maps, likelihood-based search and {\sc sparsex}. Only 17 of them have been published\footnote{Thus far, spectroscopic observations have confirmed that Reticulum\,II \citep{Koposov2015b,Simon2015,Walker2015}, Horologium\,I \citep{Koposov2015b} and Tucana\,II \citep{Walker2016} are indeed dwarf galaxies.} \citep{Bechtol2015,Drlica2015,Luque2016}. A careful reanalysis of the promising candidates detected by the {\sc sparsex} code has revealed two new candidate stellar systems in addition to those reported by \citet[][see discussion in the next section]{Drlica2015}. In this section, we briefly review {\sc sparsex}.

The {\sc sparsex} code is an overdensity detection algorithm, which is based on the matched-filter (MF) method \citep{Balbinot2011,Luque2016}. 
Briefly, we begin by binning stars into spatial pixels of $1.0\,\mathrm{arcmin}\times 1.0\,\mathrm{arcmin}$ and  colour-magnitude bins of $0.01\,\mathrm{mag}\times 0.05\,\mathrm{mag}$. We then create a grid of simple stellar populations (SSPs) with the code {\tt\sc gencmd}\footnote{https://github.com/balbinot/gencmd}. We use {\sc gencmd} along with  {\sc parsec} isochrones \citep{Bressan2012} and an initial mass function (IMF) of \citet{Kroupa2001}. We simulate several SSPs in a range of ages [$9.0\leq\log(\mathrm{Age})\leq 10.2$], metallicities ($Z=\{0.0002, 0.001, 0.007\}$) and distance ($10\leq \mathrm{D}_{\sun}\leq 200\,\mathrm{kpc}$). To account for local variations in the background CMD, we partition the sky into $10\degr\times 10\degr$ regions. We then apply {\sc sparsex} on the stellar catalogue in every sky region using the grid of the SSPs. This procedure generates one density map for each SSP model within a sky region.

\begin{figure*} 
\includegraphics[width=.85\textwidth]{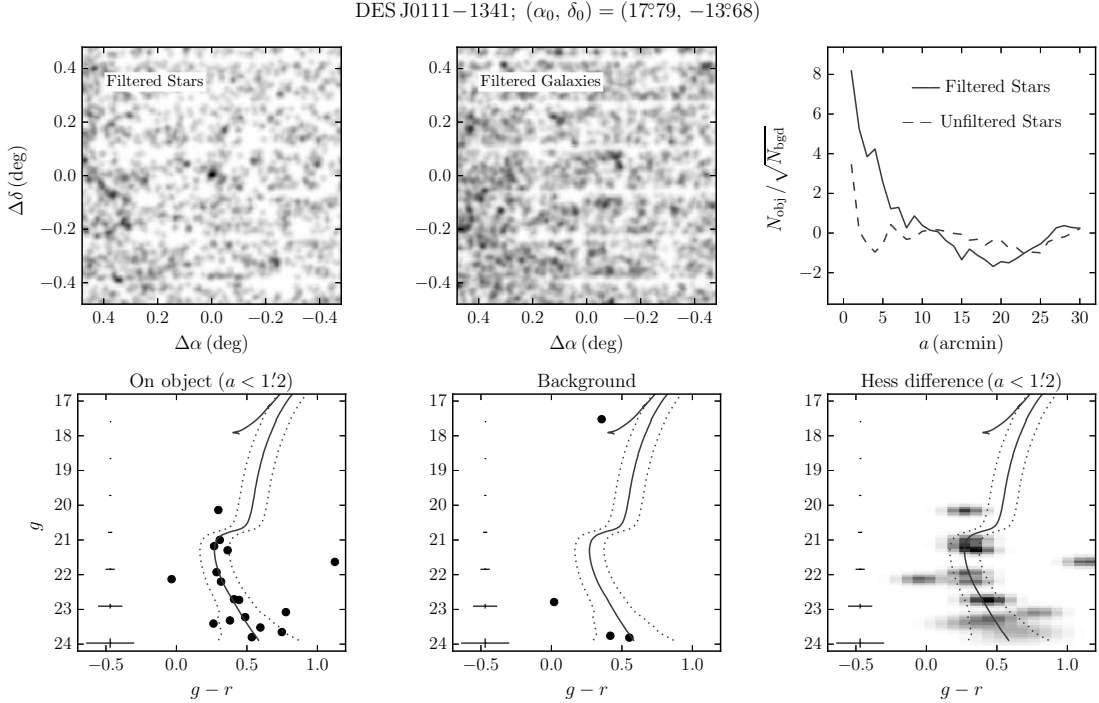}
\caption{
Top-left panel: on-sky number density map of stars around  DES\,J0111$-$1341. Only stars that lie close to the best-fitting isochrone are included. Top middle panel: similar to previous panel, but now for galaxies. Top-right panel: elliptical significance  as a function of semimajor axis $a$ from the centre of DES\,J0111$-$1341. The solid [dashed] line corresponds to isochrone-filtered [not pass the filter (unfiltered)] stars. Lower-left panel: CMD of stars within an ellipse with semimajor axis $a=2r_\mathrm{h}$ from the centre of DES\,J0111$-$1341. In this and the other two bottom panels, the best-fitting {\sc parsec} isochrone \citep{Bressan2012} is shown, along with ridge lines meant to bracket the most likely members. Lower middle panel: CMD of background stars in an elliptical annulus of equal area on the sky as the previous panel. Lower-right panel: Hess diagram of the CMD difference between stars within $a=2r_\mathrm{h}$ and background stars ($20.0\,\mathrm{arcmin}< a< 35.0\,\mathrm{arcmin}$). The mean photometric error is shown in the extreme left of each lower panel.}
\label{fig:J0111prof}
\end{figure*}

\begin{figure}\centering
\includegraphics[width=.48\textwidth]{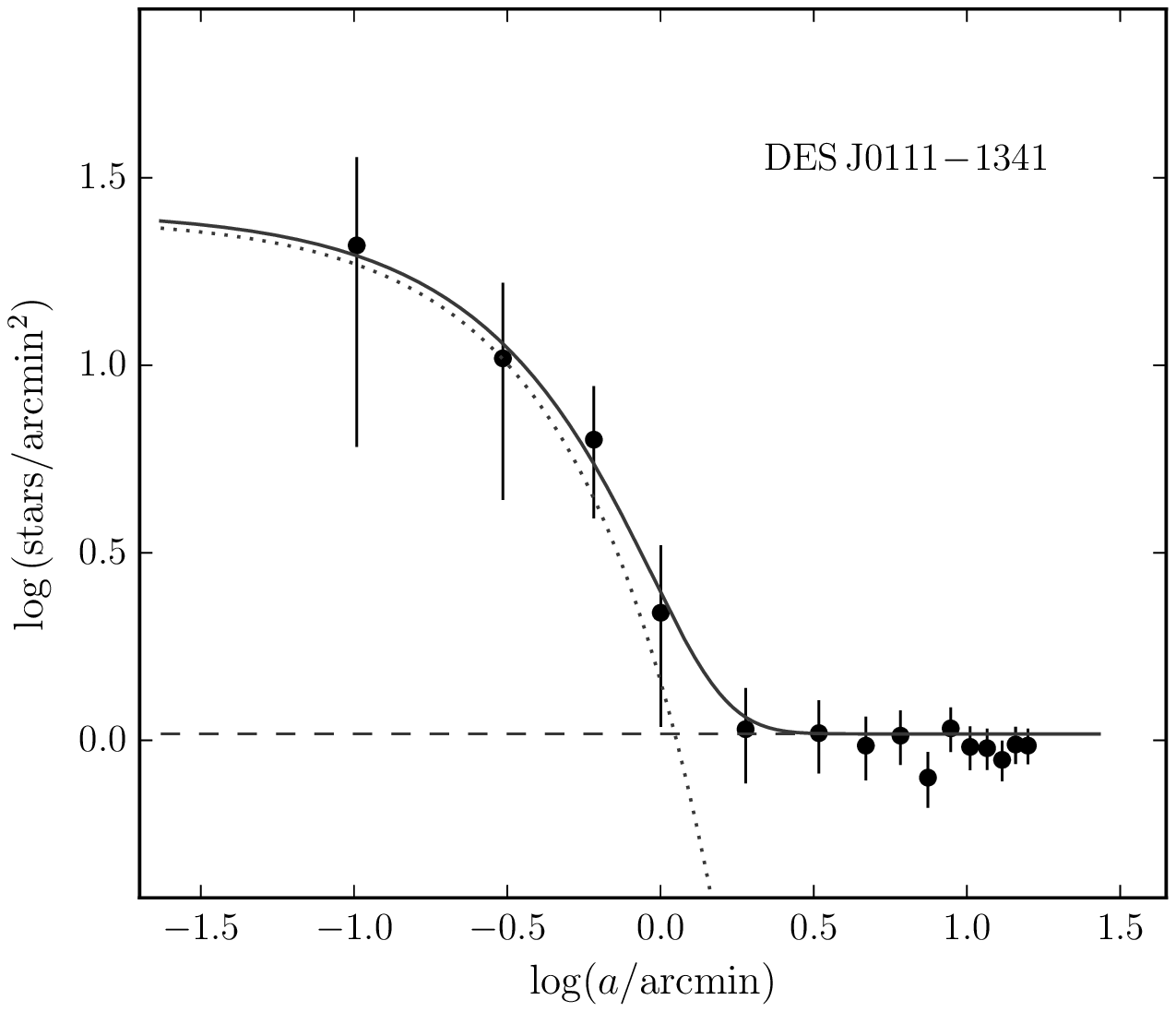}
\caption{
Solid points show a binned version of the density profile of DES\,J0111$-$1341, constructed in elliptical annuli using the derived structural parameters from the best-fitting exponential profile (see Table \ref{tab:fitpars}). The error bars are $1\sigma$ Poisson uncertainties. The dotted line represent the best-fitting of exponential profile. The horizontal dashed line shows the background level. The solid line is the combination of the background level with the exponential profile.
}
\label{fig:J0111den}
\end{figure}

To search for stellar clusters and dwarf galaxies, we convolved the set density maps with Gaussian spatial kernels of different sizes\footnote{As mentioned in \citet{Luque2016}, our range of spatial kernel sizes complements those adopted by the other two substructure search techniques. This range of kernel sizes and all possible combinations of parameters, age, metallicity and distance, allows us to detect compact objects as GCs, as well as extended objects such as dwarf galaxies.}, from $\sigma=0$ (no convolution) to $9\,\mathrm{arcmin}$. To automatically detect overdensities in each map, we use the {\tt\sc sextractor} code \citep{Bertin1996}. Finally, we selected stellar object candidates based on two criteria: (1) according to the number of times that the SSP models are detected. In this case, the 10 highest ranked candidates in each region of the sky and each convolution kernel were visually analysed. (2) According to the statistical significance of the excess number of stars  relative to background: we built a significant profile in a cumulative way, in incremental steps of $1.0\,\mathrm{arcmin}$ in radius, centred on each candidate. We then applied a simple cut in significance. All candidates with significance thresholds $> 5\sigma$  were visually analysed to discard artificial objects as well as contamination by faint galaxies \citep{Luque2016}.

Applying the method described above on DES Y2Q1 data, we successfully recovered with high significance all 19 stellar objects that have been recently reported in DES data \citep{Bechtol2015,Drlica2015,Kim22015,Kim2015b,Koposov2015,Luque2016}. 
Additionally, we detected two new candidate stellar systems potentially associated with the Sagittarius stream, DES\,J0111$-$1341 and DES\,J0225$+$0304. 
The physical properties derived for DES\,J0111$-$1341 reveal that this candidate is consistent with being an ultrafaint stellar cluster, whereas DES\,J0225$+$0304 is more consistent with being a dwarf galaxy candidate (see discussion in the next section).

\section{DES\,J0111$-$1341 and DES\,J0225$+$0304}
\label{sec:J0111_J0225}
DES\,J0111$-$1341 and DES\,J0225$+$0304 were detected with high statistical significance, $8.2\sigma$ and $7.5\sigma$, respectively. A Test Statistic (TS) for these candidates was also determined in an independent manner. The TS is based on the likelihood ratio between a hypothesis that includes an object versus a field-only hypothesis \citep[][Equation 4]{Bechtol2015}. This analysis has revealed a $\mathrm{TS\sim 15}$ ($\sim 4\sigma$) for both candidates. We do not observe an obvious overdensity of sources classified as galaxies, which reduces the possibility that the detected overdensities are caused by misclassified faint galaxies.

\begin{figure*}
\includegraphics[width=.85\textwidth]{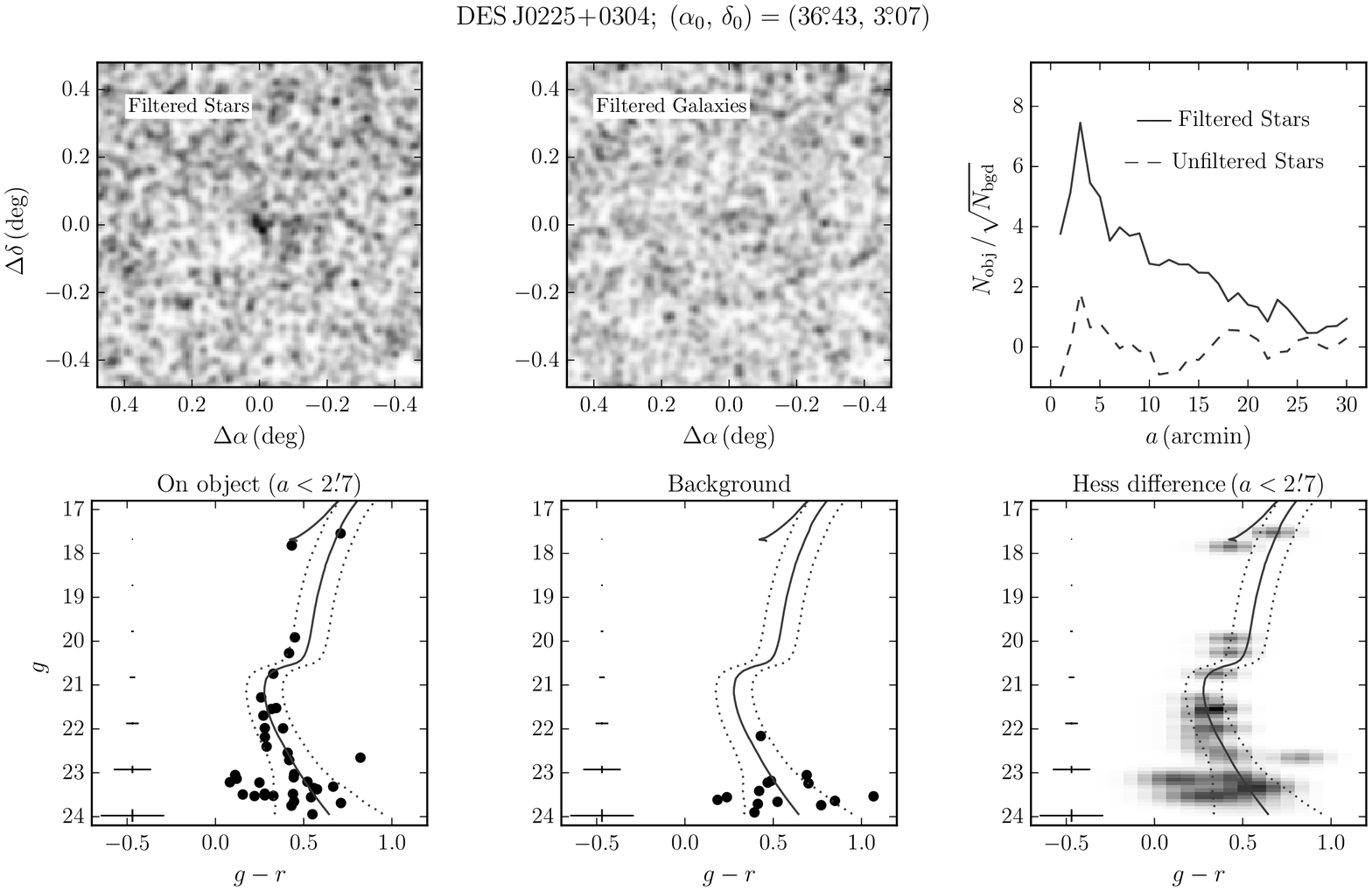}
\caption{
Top-left panel: on-sky number density map of stars around DES\,J0225$+$0304. Only stars that lie close to the best-fitting isochrone are included. Top middle panel: similar to previous panel, but now for galaxies. Top-right panel: elliptical significance as a function of semimajor axis $a$ from the centre of DES\,J0225$+$0304. The solid [dashed] line corresponds to isochrone-filtered [not pass the filter (unfiltered)] stars. Lower-left panel: CMD of stars within an ellipse with semimajor axis $a=1r_\mathrm{h}$ from the centre of DES\,J0225$+$0304. In this and the other two bottom panels, the best-fitting {\sc parsec} isochrone \citep{Bressan2012} is shown, along with ridge lines meant to bracket the most likely members. Lower-middle panel: CMD of background stars in an elliptical annulus of equal area on the sky as the previous panel. Lower-right panel: Hess diagram of the CMD difference between stars within $a=1r_\mathrm{h}$ and background stars ($25.0\,\mathrm{arcmin}< a< 45.0\,\mathrm{arcmin}$). The mean photometric error is shown in the extreme left of each lower panel.}
\label{fig:J0225prof}
\end{figure*}

\begin{figure}\centering
\includegraphics[width=.48\textwidth]{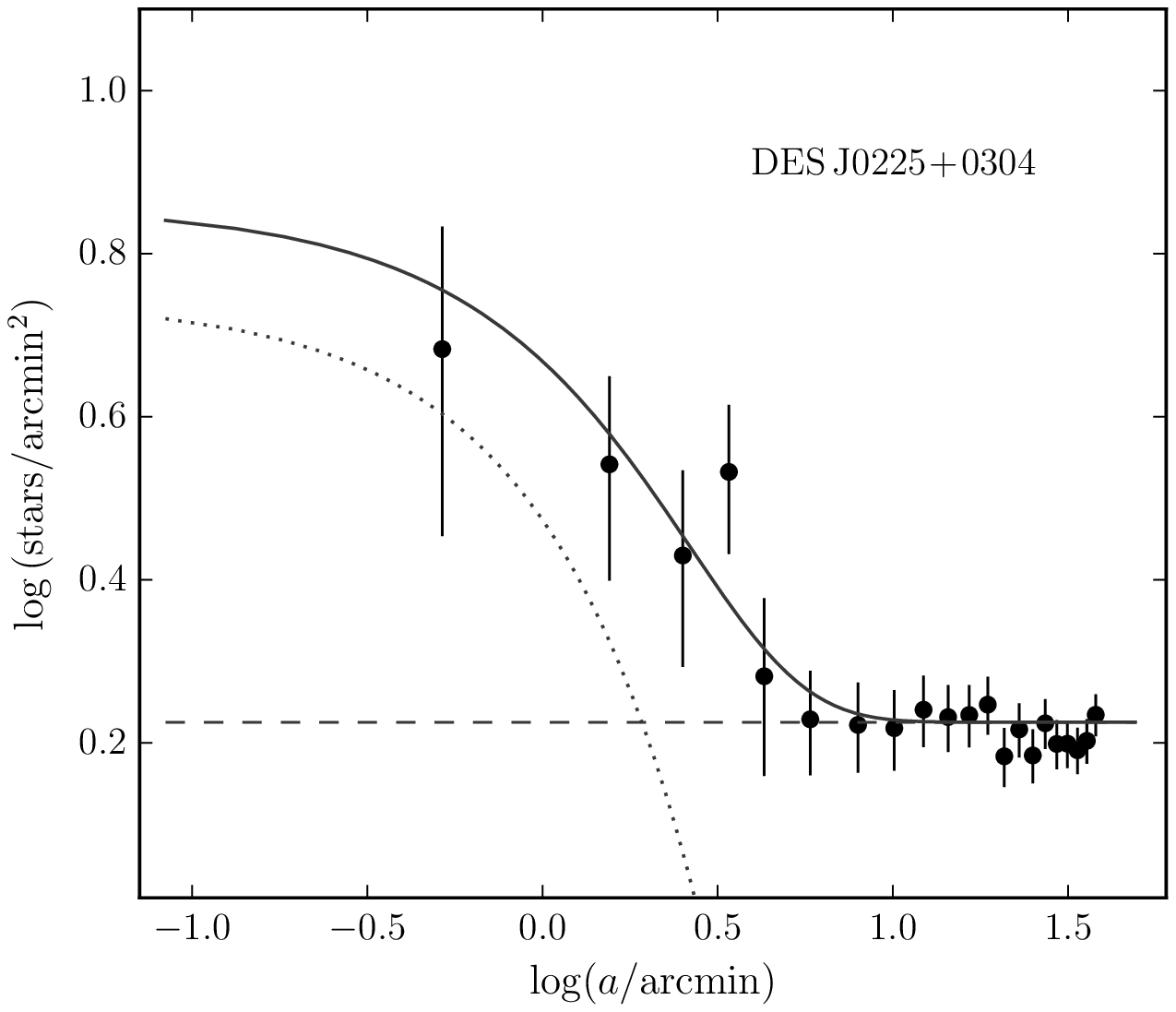}
\caption{Solid points show a binned version of the density profile of DES\,J0225$+$0304, constructed in elliptical annuli using the derived structural parameters from the best-fitting exponential profile (see Table \ref{tab:fitpars}). The error bars are $1\sigma$ Poisson uncertainties. The dotted line represent the best-fitting of exponential profile. The horizontal dashed line shows the background level. The solid line is the combination of the background level with the exponential profile.}
\label{fig:J0225den}
\end{figure}

We use the maximum likelihood technique to determine the structural and CMD parameters. To estimate the structural parameters, we assume that the spatial distribution of stars of both objects follow an exponential profile model. Following the convention of \citet{Martin2008}, we parametrize this model with six free parameters: central coordinates $\alpha_0$ and $\delta_0$, position angle $\theta$, ellipticity $\epsilon$, exponential scale radius $r_\mathrm{e}$ and background density $\Sigma_\mathrm{bgd}$. For CMD fits, we first weighted each star by the membership probability $p$ taken from the exponential density profile \citep{Pieres2016}. We then selected all the stars with a threshold of  $p\geqslant 1\%$ to fit an isochrone model. The free parameters age, $(m-M)_0$ and  $Z$ are simultaneously determined by this fitting method (for details, see \citealt{Luque2016,Pieres2016}).
To explore the parameter space, we use the {\sc emcee} module for Markov Chain Monte Carlo (MCMC; \citealt{Foreman2013})\footnote{http://dan.iel.fm/emcee/current/} sampling. We use MCMC to determine the best-fitting parameters for both the exponential profile and isochrone models. The absolute magnitudes were calculated using the prescription of \cite{Luque2016}. The inferred properties of DES\,J0111$-$1341 and DES\,J0225$+$0304 are listed in Table \ref{tab:fitpars}.

\subsection{DES\,J0111$-$1341}
\label{subsec:J0111}
DES\,J0111$-$1341 is the candidate detected with most statistical significance ($\sim 8.2\sigma$) in our sample of promising candidates. In the top-left panel of Fig. \ref{fig:J0111prof}, we show the density map constructed using stars inside the isochrone filter. For comparison, we show in the top middle panel the density map of objects classified as galaxies. Note the prominent  stellar overdensity centred on DES\,J0111$-$1341. The top-right panel shows the elliptical significance profile. It is defined as the ratio of the number of stars inside a given ellipse and in excess of the background ($N_\mathrm{bgd}$), $N_\mathrm{obj}$, to the expected fluctuation in the same background, i.e, $N_\mathrm{obj}/\sqrt{N_\mathrm{bgd}}$. $N_\mathrm{obj}=(N_\mathrm{obs}-N_\mathrm{bgd})$, where $N_\mathrm{obs}$ is the total number of observed stars. We build the elliptical significance profile using cumulative ellipses with semimajor axis $a$ centred on the object. $N_\mathrm{bgd}$ is computed within an elliptical annulus at $30\,\mathrm{arcmin} <a<34\,\mathrm{arcmin}$ from DES\,J0111$-$1341 \citep{Luque2016}. Note that the higher peak of significance (PS) is clearly steeper for the filtered stars according to our best-fitting isochrone model. In the same figure, the CMD for DES\,J0111$-$1341 is shown in the bottom-left panel. Only stars inside an ellipse with semimajor axis $a=2r_\mathrm{h}$ are shown. The CMD shows predominantly MS stars. The bottom middle panel shows the CMD of background stars contained within an elliptical annulus of equal area as the previous panel. In both CMDs, we show the filter based on our best-fitting isochrone (see \citealt{Luque2016}). The Hess difference between the stars inside an ellipse with semimajor axis $a=2r_\mathrm{h}$ and background stars ($20.0\,\mathrm{arcmin} < a < 35.0\,\mathrm{arcmin}$), this latter scaled to the same area, is shown in the bottom right-panel. In Fig. \ref{fig:J0111den}, we show the binned stellar density profile for DES\,J0111$-$1341. The best-fitting exponential model is also overplotted. In both cases, we took into account the ellipticity of the object.

The physical size ($r_\mathrm{h}\sim 4.55\,\mathrm{pc}$) of DES\,J0111$-$1341 is comparable with the size of GCs associated with the Sagittarius stream [e.g., Terzan\,7 ($r_\mathrm{h}\sim 5.10\,\mathrm{pc}$) and NGC\,6715 ($r_\mathrm{h}\sim 6.3\,\mathrm{pc}$); \citealt{Forbes2010,Harris2010,LawMajewski2010}]. However, its luminosity ($M_V\sim +0.3$) is inconsistent with this class of objects [$M_V \sim -5$ (Terzan\,7) and $M_V\sim -10$ (NGC\,6715); \citealt{Forbes2010,Harris2010,LawMajewski2010}]. Therefore, its low luminosity and small size place DES\,J0111$-$1341 among the MW ultrafaint stellar clusters (see size--luminosity plane, Fig. \ref{fig:magrh}). In particular, its luminosity is comparable to Kim\,1 ($M_V\sim +0.3$; \citealt{Kim2015a}). However, DES\,J0111$-$1341 is fainter than DES\,1 \citep{Luque2016}, Koposov\,1, Koposov\,2 \citep{Koposov2007} and Mu\~noz\,1 \citep{Munoz2012}.

\subsection{DES\,J0225$+$0304}
\label{subsec:J0225}
\begin{figure}\centering
\includegraphics[width=.49\textwidth]{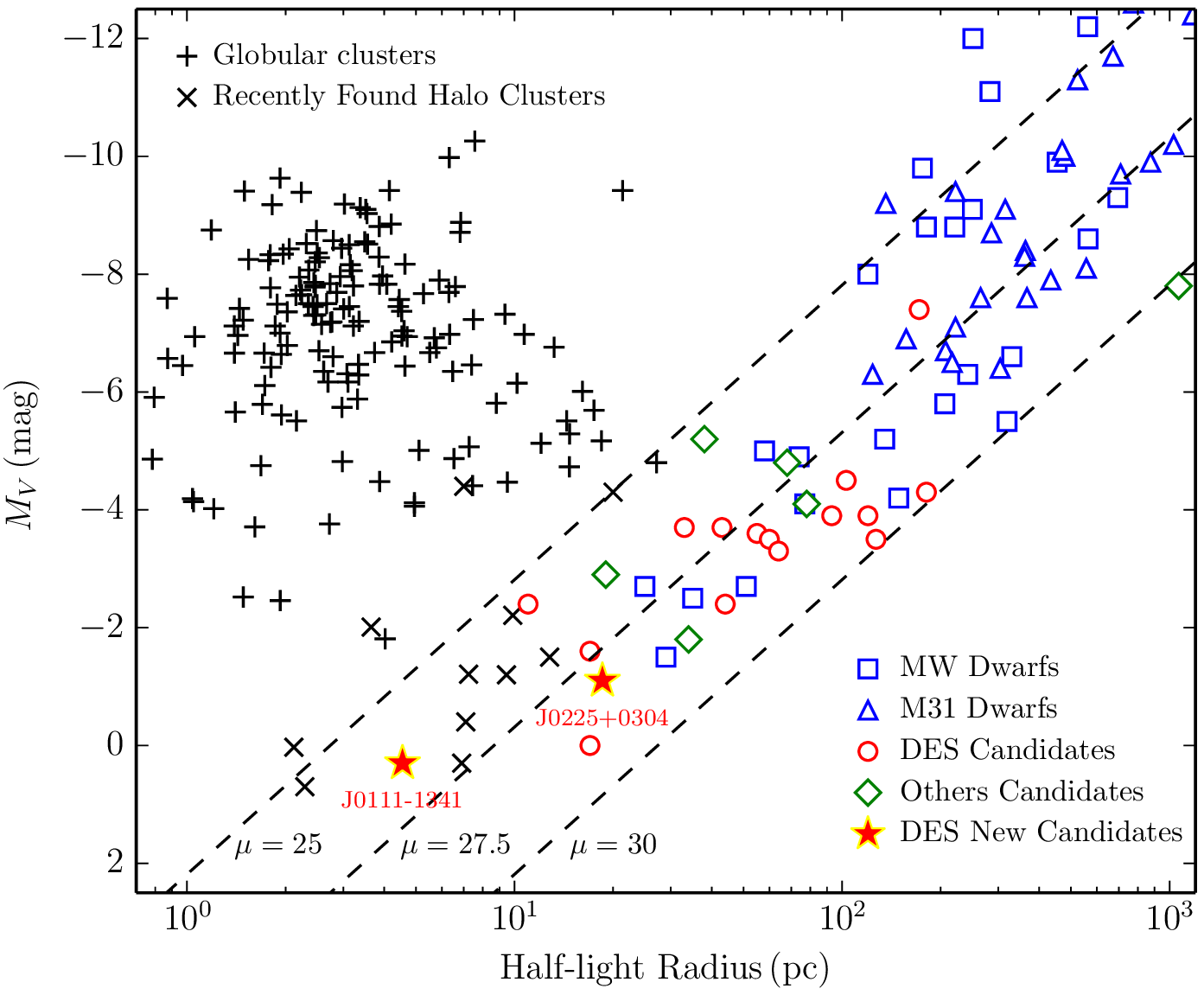}
\caption{Absolute magnitude as a function of half-light radius. MW GCs (`$+$' symbols; \citealt{Harris2010}), outer halo clusters with ambiguous classification (`$\times$' symbols; \citealt{Koposov2007}; \citealt{Belokurov2010}; \citealt{Munoz2012}; \citealt{Balbinot2013};  \citealt{Laevens2014,Laevens2015a}; \citealt{Kim2015a,Kim22015,Kim32016}; \citealt{Luque2016}), MW dwarf galaxies (blue squares; \citealt{McConnachie2012}), M\,31 dwarf galaxies (blue triangles; \citealt{McConnachie2012}), previously reported dwarf galaxy candidates in the DES footprint (red circles; \citealt{Bechtol2015}; \citealt{Drlica2015}), other recently reported dwarf galaxies (green diamond; \citealt{Kim2015a};  \citealt{Laevens2015b,Laevens2015a}; \citealt{Martin2015a}; \citealt{Torrealba2016}), and DES new candidates (red stars) are shown. Note that DES\,J0111$-$1341 clearly lies inside the region inhabited  by ultrafaint stellar clusters, whereas DES\,J0225$+$0304 occupies the ambiguous region between stellar clusters and dwarf galaxies. The dashed lines indicate contours of constant surface brightness at $\mu=\{25,\, 27.5,\, 30\}\,\mathrm{mag\,arsec^{-2}}$.}\label{fig:magrh}
\end{figure}

Figs \ref{fig:J0225prof} and \ref{fig:J0225den} show the analogous information as Figs \ref{fig:J0111prof} and \ref{fig:J0111den} for DES\,J0225+0304. The physical size ($r_\mathrm{h}\sim 18.55\,\mathrm{pc}$) and luminosity ($M_V\sim -1.1$) place it 
in an ambiguous region of size--luminosity space between stellar clusters and dwarf
galaxies (see Fig. \ref{fig:magrh}). DES\,J0225$+$0304 is elongated ($\epsilon\sim 0.61$) and has a physical size similar to an extended GC or a very small faint dwarf galaxy. In fact, the physical size, luminosity and ellipticity of DES\,J0225$+$0304 are comparable to the properties of the Tucana\,V stellar system ($r_\mathrm{h}\simeq 17\,\mathrm{pc}$, $M_V\simeq -1.6$ and $\epsilon\simeq 0.7$; \citealt{Drlica2015}). 

\subsection{Association with the Sagittarius stream}
As mentioned in Section \ref{sec:Sgt}, DES\,J0111$-$1341 and DES\,J0225$+$0304  are probably associated with the Sagittarius dwarf stream. Their $\log(\mathrm{Age})$, $\mathrm{[Fe/H]}$ and $\mathrm{D}_{\sun}$ (see Table \ref{tab:fitpars}) are well bracketed by the age, metallicity and distance ranges determined in Section \ref{sec:spread_met} for the stream. In fact, the inferred ages and metallicities are very similar for both DES\,J0111$-$1341 and DES\,J0225$+$0304, and agree very well with the isochrone fit to the mean RGB colours of the stream, $\log(\mathrm{Age})=10.02$, $\mathrm{[Fe/H]}=-1.34$ and $\mathrm{D}_{\sun}=24.5\,\mathrm{kpc}$. 

To better explore this association, we estimate the distance of the two new candidates to the Sagittarius orbital plane ($D_\mathrm{orb}$). For this purpose, we use the best-fitting Sagittarius orbital plane\footnote{The best-fitting plane was performed by using M-giant stars detected in 2MASS data (for details, see \citealt{Majewski2003}).} determined by \citet{Majewski2003}. We then obtain a distance of $\sim 1.73$ and $\sim 0.50\,\mathrm{kpc}$ for DES\,J0111$-$1341 and DES\,J0225$+$0304, respectively. When we compare the $D_\mathrm{orb}$ of the new candidates with the $D_\mathrm{orb}$ determined for GCs  associated with the Sagittarius dwarf \citep[and references therein]{Bellazzini2003}, we note that DES\,J0111$-$1341 has a  $D_\mathrm{orb}$ similar to Terzan\,7 ($\sim 1.89\,\mathrm{kpc}$), whereas that the $D_\mathrm{orb}$ of DES\,J0225$+$0304 is comparable to NGC\,6715 ($\sim 0.45\,\mathrm{kpc}$). These results indicate that both DES\,J0111$-$1341 and DES\,J0225$+$0304 are very close indeed to the Sagittarius plane, something that strongly increases the likelihood of their association with the Sagittarius stream. However, there are GCs spatially compatible with the orbit of Sagittarius (e.g., NCG\,4147 and NGC\,288) but not associated with Sagittarius when their radial velocities and proper motions are considered \citep{LawMajewski2010}. This suggests that the spectroscopic determination of the radial velocity, and the proper motion of these systems are both crucial to confirm that association.

We use a random sampling technique to give a statistical argument for this possible association. For this purpose, we use  the sample of known  star clusters and dwarf galaxies from various recent sources  \citep{Harris2010,McConnachie2012,Balbinot2013,Laevens2014,Laevens2015a,Laevens2015b,Bechtol2015,Drlica2015,Koposov2015,Kim2015a,Kim22015,Kim32016,Martin2015a,Luque2016}. The null hypothesis assumes that the stellar systems from our sample are not associated with the Sagittarius dwarf galaxy, thus we removed the four GCs confirmed to be associated with the Sagittarius dwarf (NGC\,6715, Arp\,2, Terzan\,7, Terzan\,8). First we calculate $D_\mathrm{orb}$ for each stellar system. We then randomly selected two systems from the sample, assigning an equal selection probability to each system. After performing  $10^6$ selections, we estimated a $0.08$ probability of finding two stellar systems with $D_\mathrm{orb} \leqslant 1.73\,\mathrm{kpc}$. While it is true that this probability value is not negligible, these randomly drawn pairs of objects are not necessarily as close to the Sagittarius orbit as our candidates.
 
\section{Conclusions}
\label{sec:conclusions}
In this paper, we report the discovery of two new candidate stellar systems in the constellation of Cetus using  DES Y2Q1 data. These objects add to the 19 star systems that have been found in the first 2\,yr of DES \citep{Bechtol2015,Drlica2015,Kim2015b,Koposov2015,Luque2016}. DES\,J0111$-$1341 is a compact ($r_\mathrm{h}\sim 4.55\,\mathrm{pc}$) and ultrafaint ($M_V\sim +0.3$) stellar cluster, whereas DES\,J0225$+$0304 in faint ($M_V\sim -1.1$) and has a physical size ($r_\mathrm{h}\sim 18.55\,\mathrm{pc}$) comparable to a very small faint dwarf galaxy. These new stellar systems appear to be at a heliocentric distance $\mathrm{D}_{\sun}\sim 25\,\mathrm{kpc}$. 

There are several lines of evidence that suggest that our new candidates are associated with the Sagittarius stream: (i) they lie on the edges of the Sagittarius stream, as can be seen in Fig. \ref{fig:maps_Sg} (red circles). (ii) The CMD parameters (age, metallicity and distance) determined for these new candidates lie within the metallicity and age range determined for the Sagittarius stream using the same DES data (Section \ref{sec:spread_met}). In particular, they are consistent with the parameters inferred by fitting the mean CMD locus of the stream stars. (iii) The distances $D_\mathrm{orb}$ of our candidates to the Sagittarius orbital plane, $\sim 1.73\,\mathrm{kpc}$ (DES\,J0111$-$1341) and $\sim 0.50\,\mathrm{kpc}$ (DES\,J0225$+$0304), are comparable to GCs previously associated with the Sagittarius dwarf, more specifically Terzan\,7 and NGC\,6715 \citep{Bellazzini2003}. Therefore, we speculate that these candidates are likely associated with the Sagittarius stream. However, the spectroscopic determination of the radial velocity and proper motion of these substructures will be very useful to confirm this hypothesis. Furthermore, the dynamic mass, derived from the velocity dispersion, will help to confirm the nature of our candidates. If all of our hypotheses are confirmed, DES\,J0225$+$0304 would be the first ultrafaint dwarf galaxy associated with the Sagittarius dwarf stream. It would also be the first confirmed case of an ultrafaint satellite of a satellite. 

\begin{center}
\begin{table}
\caption{Properties of DES\,J0111$-$1341 and DES\,J0225$+$0304}
\hspace*{-.6cm}
\scalebox{.95}{
\begin{tabular}{lccc}\hline
Parameters & DES\,J0111$-$1341 & DES\,J0225$+$0304 &Unit\\\hline
$\alpha_0\,(J2000)$& $\mathrm{01\!:\!11\!:\!10.3^{+0.40}_{-0.48}}$ &   $\mathrm{02\!:\!25\!:\!42.4^{+1.52}_{-1.60}}$ &$\mathrm{h\!:\!m\!:\!s}$\\
$\delta_0\,(J2000)$& $\mathrm{-13\!:\!41\!:\!05.4^{+5.4}_{-6.6}}$ & $\mathrm{03\!:\!04\!:\!10.1^{+45.6}_{-39.6}}$ &$\degr\!:\!\arcmin\!:\!\arcsec$\\
$l$&$142.83$&$163.58$&$\mathrm{deg}$\\
$b$&$-75.79$&$-52.20$&$\mathrm{deg}$\\
$\Lambda$&$86.61$&$111.02$&$\mathrm{deg}$\\
$B$&$-3.97$&$1.24$&$\mathrm{deg}$\\
$\theta$ & $-53.24^{+31.70}_{-23.24}$ & $31.25^{+11.48}_{-13.39}$ & $\mathrm{deg}$\\
$\epsilon$ & $0.27^{+0.20}_{-0.17}$ &$0.61^{+0.14}_{-0.23}$&\\
$\Sigma_{\mathrm{bgd}}$ & $1.040^{+0.001}_{-0.001}$ & $1.679^{+0.002}_{-0.002}$ &$\tfrac{\mathrm{stars}}{\mathrm{arcmin}^2}$\\
$\mathrm{D}_\odot$ & $26.5^{+1.3}_{-1.3}$ & $23.8^{+0.7}_{-0.5}$ & $\mathrm{kpc}$\\
$r_\mathrm{h}$\textsuperscript{$a$} & $0.59^{+0.17}_{-0.12}$ & $2.68^{+1.33}_{-0.70}$&$\mathrm{arcmin}$\\
$r_\mathrm{h}$ &$4.55^{+1.33}_{-0.95}$\textsuperscript{$b$} & $18.55^{+9.22}_{-4.86}$\textsuperscript{$c$} &$\mathrm{pc}$\\
$M_V$ & $+0.3^{+0.9}_{-0.6}$ & $-1.1^{+0.5}_{-0.3}$ &$\mathrm{mag}$\\
$D_{\mathrm{orb}}$&$\sim 1.73$&$\sim 0.50$&$\mathrm{kpc}$\\
$\mathrm{[Fe/H]}$\textsuperscript{$d$} & $-1.38^{+0.07}_{-0.05}$ &$-1.26^{+0.03}_{-0.03}$&\\
$\log(\mathrm{Age})$ & $10.06^{+0.02}_{-0.02}$ &$10.07^{+0.01}_{-0.01}$&\\
$(m-M)_0$ & $17.12^{+0.11}_{-0.11}$ &$16.88^{+0.06}_{-0.05}$&\\\hline
\end{tabular}}\label{tab:fitpars}\\[.1cm]
\leftline{\scalebox{.95}{\textit{Notes}. \textsuperscript{$a$}\footnotesize{Using the relation, $r_\mathrm{h}=1.68 r_\mathrm{e}$ \citep{Martin2008}.}}}
\leftline{\scalebox{.95}{\textsuperscript{$b$}\footnotesize{Adopting a distance of $26.5\,\mathrm{kpc}$.}}} 
\leftline{\scalebox{.92}{\textsuperscript{$c$}\footnotesize{Adopting a distance of $23.8\,\mathrm{kpc}$.}}}
\leftline{\scalebox{.95}{\textsuperscript{$d$}\footnotesize{Adopting $\mathrm{Z}_{\sun} = 0.0152$ \citep{Bressan2012}}.}}
\end{table} 
\end{center}

As for the properties of the stream itself, the star count histograms constructed across the Sagittarius stream show a possible excess of stars at $B\sim 8\degr$. However, this putative excess is only clearly visible when we use bin sizes of $0\fdg 6 \lesssim \Delta B \lesssim 0\fdg 7$. Therefore, we do not claim a detection of the branching of the stream. We found no further direct evidence of additional stream  substructures to those already known to exist.

Finally, decontaminated Hess diagrams of the  Sagittarius stream allowed us to determine a metallicity spread ($-2.18 \lesssim\mathrm{[Fe/H]} \lesssim -0.95$) as well as a distance gradient ($23\,\mathrm{kpc} \lesssim \mathrm{D}_{\sun} \lesssim 29\,\mathrm{kpc}$). This suggests that the stream is composed of more than one stellar population. Our determination of distance gradient is consistent with those determined by \citet{Koposov2012}. However, metallicity determinations in the literature suggest that the stream in the celestial equator contains more metal-rich stars than those determined in this work \citep[see, e.g.,][]{Koposov2012,deBoer2015}.

In the future, DES will acquire additional imaging data in this region, allowing even more significant studies of the region in which the Sagittarius stream crosses the equator.

\section*{Acknowledgements} 
This paper has gone through internal review by the DES collaboration.

Funding for the DES Projects has been provided by the U.S. Department of Energy, the U.S. National Science Foundation, the Ministry of Science and Education of Spain, the Science and Technology Facilities Council of the United Kingdom, the Higher Education Funding Council for England, the National Center for Supercomputing Applications at the University of Illinois at Urbana-Champaign, the Kavli Institute of Cosmological Physics at the University of Chicago, the Center for Cosmology and Astro-Particle Physics at the Ohio State University, the Mitchell Institute for Fundamental Physics and Astronomy at Texas A\&M University, Financiadora de Estudos e Projetos, Funda{\c c}{\~a}o Carlos Chagas Filho de Amparo \`a Pesquisa do Estado do Rio de Janeiro, Conselho Nacional de Desenvolvimento Cient{\'i}fico e Tecnol{\'o}gico and the Minist{\'e}rio da Ci{\^e}ncia, Tecnologia e Inova{\c c}{\~a}o, the Deutsche Forschungsgemeinschaft and the Collaborating Institutions in the Dark Energy Survey.  The DES data management system is supported by the National Science Foundation under Grant Number AST-1138766. The DES participants from Spanish institutions are partially supported by MINECO under grants AYA2012-39559, ESP2013-48274, FPA2013-47986, and Centro de Excelencia Severo Ochoa SEV-2012-0234, some of which include ERDF funds from the European Union.

The Collaborating Institutions are Argonne National Laboratory, the University of California at Santa Cruz, the University of Cambridge, Centro de Investigaciones En{\'e}rgeticas, Medioambientales y Tecnol{\'o}gicas-Madrid, the University of Chicago, University College London, the DES-Brazil Consortium, the University of Edinburgh, the Eidgen{\"o}ssische Technische Hochschule (ETH) Z{\"u}rich,  Fermi National Accelerator Laboratory, the University of Illinois at Urbana-Champaign, the Institut de Ci\`encies de l'Espai (IEEC/CSIC), the Institut de F{\'i}sica d'Altes Energies, Lawrence Berkeley National Laboratory, the Ludwig-Maximilians Universit{\"a}t M{\"u}nchen and the associated Excellence Cluster Universe, the University of Michigan, the National Optical Astronomy Observatory, the University of Nottingham, The Ohio State University, the University of Pennsylvania, the University of Portsmouth, SLAC National Accelerator Laboratory, Stanford University, the University of Sussex, and Texas A\&M University.

The DES data management system is supported by the National Science Foundation under Grant Number AST-1138766. The DES participants from Spanish institutions are partially supported by MINECO under grants AYA2012-39559, ESP2013-48274, FPA2013-47986, and Centro de Excelencia Severo Ochoa SEV-2012-0234.

Research leading to these results has received funding from the European Research Council under the European Union{'}s Seventh Framework Programme (FP7/2007-2013) including ERC grant agreements 240672, 291329, and 306478.

EB acknowledges financial support from the European Research Council (ERC-StG-335936, CLUSTERS).




\bibliographystyle{mnras}
\bibliography{bib} 
\vspace*{1.3em}





{\small\it\noindent
$^1$Instituto de F\'\i sica, UFRGS, Caixa Postal 15051, Porto Alegre, RS - 91501-970, Brazil\\
$^2$Laborat\'orio Interinstitucional de e-Astronomia - LIneA, Rua Gal. Jos\'e Cristino 77, Rio de Janeiro, RJ - 20921-400, Brazil\\
$^3$Fermi National Accelerator Laboratory, P. O. Box 500, Batavia, IL 60510, USA\\
$^4$Cerro Tololo Inter-American Observatory, National Optical Astronomy Observatory, Casilla 603, La Serena, Chile\\
$^5$National Center for Supercomputing Applications, 1205 West Clark St., Urbana, IL 61801, USA\\
$^6$Department of Physics, University of Surrey, Guildford GU2 7XH, UK\\
$^7$George P. and Cynthia Woods Mitchell Institute for Fundamental Physics and Astronomy, and Department of Physics and Astronomy, Texas A\&M University, College Station, TX 77843,  USA\\
$^{8}$Observat\'orio Nacional, Rua Gal. Jos\'e Cristino 77, Rio de Janeiro, RJ - 20921-400, Brazil\\
$^{9}$Kavli Institute for Cosmological Physics, University of Chicago, Chicago, IL 60637, USA\\
$^{10}$Lawrence Berkeley National Laboratory, 1 Cyclotron Road, Berkeley, CA 94720, USA\\
$^{11}$Department of Physics and Astronomy, University of Pennsylvania, Philadelphia, PA 19104, USA\\ 
$^{12}$Department of Physics \& Astronomy, University College London, Gower Street, London, WC1E 6BT, UK\\
$^{13}$Department of Physics and Electronics, Rhodes University, PO Box 94, Grahamstown, 6140, South Africa\\
$^{14}$CNRS, UMR 7095, Institut d'Astrophysique de Paris, F-75014, Paris, France\\
$^{15}$Sorbonne Universit\'es, UPMC Univ Paris 06, UMR 7095, Institut d'Astrophysique de Paris, F-75014, Paris, France\\
$^{16}$Kavli Institute for Particle Astrophysics \& Cosmology, P. O. Box 2450, Stanford University, Stanford, CA 94305, USA\\
$^{17}$SLAC National Accelerator Laboratory, Menlo Park, CA 94025, USA\\
$^{18}$Department of Astronomy, University of Illinois, 1002 W. Green Street, Urbana, IL 61801, USA\\
$^{19}$Institut de Ci\`encies de l'Espai, IEEC-CSIC, Campus UAB, Carrer de Can Magrans, s/n,  08193 Bellaterra, Barcelona, Spain\\
$^{20}$Institut de F\'{\i}sica d'Altes Energies (IFAE), The Barcelona Institute of Science and Technology, Campus UAB, 08193 Bellaterra (Barcelona) Spain\\
$^{21}$Institute of Cosmology \& Gravitation, University of Portsmouth, Portsmouth, PO1 3FX, UK\\
$^{22}$School of Physics and Astronomy, University of Southampton,  Southampton, SO17 1BJ, UK\\
$^{23}$Department of Physics, IIT Hyderabad, Kandi, Telangana 502285, India\\
$^{24}$Department of Astronomy, University of Michigan, Ann Arbor, MI 48109, USA\\
$^{25}$Department of Physics, University of Michigan, Ann Arbor, MI 48109, USA\\
$^{26}$Department of Astronomy, University of California, Berkeley,  501 Campbell Hall, Berkeley, CA 94720, USA\\
$^{27}$Australian Astronomical Observatory, North Ryde, NSW 2113, Australia\\
$^{28}$Center for Cosmology and Astro-Particle Physics, The Ohio State University, Columbus, OH 43210, USA\\
$^{29}$Department of Astronomy, The Ohio State University, Columbus, OH 43210, USA\\
$^{30}$Instituci\'o Catalana de Recerca i Estudis Avan\c{c}ats, E-08010 Barcelona, Spain\\
$^{31}$Jet Propulsion Laboratory, California Institute of Technology, 4800 Oak Grove Dr., Pasadena, CA 91109, USA\\
$^{32}$Department of Physics and Astronomy, Pevensey Building, University of Sussex, Brighton, BN1 9QH, UK\\
$^{33}$Centro de Investigaciones Energ\'eticas, Medioambientales y Tecnol\'ogicas (CIEMAT), Madrid, Spain\\
$^{34}$Universidade Federal do ABC, Centro de Ci\^encias Naturais e Humanas, Av. dos Estados, 5001, Santo Andr\'e, SP, Brazil, 09210-580\\
$^{35}$Computer Science and Mathematics Division, Oak Ridge National Laboratory, Oak Ridge, TN 37831
}

\bsp	
\label{lastpage}
\end{document}